\documentclass[11pt]{article}

\usepackage[utf8]{inputenc}
\usepackage[T1]{fontenc}
\usepackage[margin=1in]{geometry}
\usepackage{times}
\usepackage{microtype}
\usepackage{booktabs}
\usepackage{hyperref}
\usepackage{xurl}
\usepackage{xcolor}
\usepackage{colortbl}
\usepackage{enumitem}
\usepackage[numbers,sort&compress]{natbib}
\usepackage{tikz}
\usetikzlibrary{positioning,arrows.meta,calc}
\usepackage{orcidlink}

\hypersetup{
    colorlinks=true,
    linkcolor=blue,
    citecolor=blue,
    urlcolor=blue
}

\title{The Research Guide: From Informal Role to Profession}

\author{
    Sergey V. Samsonau\,\orcidlink{0000-0002-0835-2970} and Matthew Pearce\,\orcidlink{0009-0000-1881-4561}\\[6pt]
    Authentic Research Partners\\
    Princeton, NJ
}

\date{April 2026}

\begin{document}
\maketitle

% ============================================================================
% Abstract
% ============================================================================

\begin{abstract}
Guiding others through authentic scientific research outside of PhD programs has been practiced for decades in specialized secondary schools, undergraduate research programs, and independent settings. These practitioners work in the middle, between the classroom science teacher and the PhD advisor, guiding learners with aptitude or serious interest. Sport and music have dedicated professions for this middle position (the school-team coach and the school band director); research does not. This paper names that missing profession the \textbf{Research Guide}: the practitioner who develops another person's capacity to do research, from framing a question to communicating findings.

Hundreds of thousands of middle and high school students already pursue authentic research each year, even more college undergraduates participate in research with a faculty member, and millions of adults engage in citizen science. In current practice, the programs that serve this middle group mostly default to a simplified version of the PhD apprenticeship model structured around one mentor with a few students at a time, without systematic training; they overwhelmingly frame research as the hypothetico-deductive cycle alone.

The role calls for \emph{cognitive apprenticeship}, a pedagogical approach in which an expert's tacit moves on open-ended problems are made visible and scaffolded, then faded as the learner develops, while the research outcomes themselves remain unpredictable. It spans multiple modes of inquiry (not only the hypothetico-deductive cycle) and demands a combination that no existing training program produces: pedagogy, research methodology, developmental assessment, risk and productive struggle management, domain flexibility, and community building. Together these demands warrant a dedicated profession: a named role, a training pathway, a career ladder, hiring standards, and institutional recognition.
\end{abstract}


% ============================================================================
% 1. Introduction
% ============================================================================

\section{Introduction}
\label{sec:intro}

At Bergen County Academies in Hackensack, New Jersey, a team of staff hold the title ``Research Teacher,'' a term the school appears to have coined internally.\footnote{Bergen County Academies, Research Teachers; \url{https://www.bergen.org/domain/712}.} At Stuyvesant High School in Manhattan, the role is titled ``Research Coordinator,'' held by a biology teacher who runs the school's research program.\footnote{Stuyvesant High School, Talos blog, ``Research Opportunities $-$ Links from Research Coordinator, Mr. Econome!''; \url{https://talos.stuy.edu/cms/pages/stuyvesant-blog/research-opportunities-links-research-coordinator-mr-econome/}.} At Thomas Jefferson High School for Science and Technology in Fairfax, Virginia, one of the most prestigious STEM high schools in the country, there are specialized research teachers called ``Lab Technology Teachers'' and other science and technology faculty may also mentor students.\footnote{Thomas Jefferson High School for Science and Technology, Senior Research Labs; \url{https://tjhsst.fcps.edu/academics/research-program}.} Three schools, three titles, no consistency.

This pattern repeats at the university level. Princeton has a ``Director of the Office of Undergraduate Research.'' Indiana University posted a ``Director of Undergraduate Research and Mentor Development.'' Many universities employ ``Undergraduate Research Coordinators,'' but these are typically administrative roles managing budgets and compliance, not mentoring roles. The actual mentoring is done by faculty absorbing the work into already full portfolios, or by graduate students with no mentoring training.

Compare this with any established profession. Every school calls a math teacher a math teacher. Every hospital calls a nurse a nurse. The standardization of titles is not only for bureaucratic convenience; it is the foundation on which training programs, certification, career tracks, and professional organizations are built. When a role has no name, none of these can follow.

This paper proposes that the person who guides others through authentic scientific research, whom we call the \textbf{Research Guide}, occupies a distinct professional role that has not been formally recognized. The role requires a specific combination of skills that no existing training program produces. The absence of this profession has concrete consequences: research mentoring depends on individual dedication, programs are fragile with single points of failure, and access is stratified by wealth rather than talent. The profession spans every level where research is learned before and alongside the doctoral track: middle and high school students, undergraduates, and adults outside academia (domain experts, career changers, citizen and participatory scientists); this paper addresses these populations.

The claim is not that nobody does this work; people do it every day, across every level of education, in every country. But outside of doctoral training, where apprenticeship is the appropriate model, most do this work in the same apprenticeship style, one practitioner with a few students at a time, without the structured pedagogy that would let them reach students at classroom scale. The work exists; the reach, the consistency, and the instructor-to-student ratio that a trained profession could sustain do not. They are, in the language of a parallel movement, ``a hidden yet fundamental group'' \cite{brett2017}.

The role is not new. It has been practiced, without a name, in some of the most successful science education systems in history. In 1963, the Soviet government formally established specialized \emph{fizmat} (physics-mathematics) boarding schools affiliated with major universities, with curricula designed by leading scientists including Isaak Kikoin and Andrey Kolmogorov \cite{gerovitch2019}. These schools placed working researchers in direct contact with talented secondary students through problem sets that externalized expert reasoning, and institute-grade laboratories where the research process was practiced rather than described: cognitive apprenticeship (Section~\ref{sec:tiers}) in everything but the name. In England, the Nuffield Advanced Physics project (1960s--70s) replaced rote instruction with inquiry-based learning, emphasizing ``learning through doing'' and scientific habits of thought \cite{nuffield2010}. Both traditions produced extraordinary outcomes (Fields Medalists, Nobel laureates, generations of scientists) and both depended entirely on exceptional individuals for whom no professional category existed.

The International Science and Engineering Fair (ISEF), the world's largest pre-college science competition, demonstrates both the demand and the gap: approximately 175,000 students compete in affiliated fairs annually,\footnote{Society for Science press release, 2024; \url{https://www.societyforscience.org/press-release/columbus-to-host-regeneron-isef-2025/}.} but ISEF is a competition, not a mentoring system. Its structure requires a well-formulated research question before a project begins, which assumes the question has already been formulated. This ignores the discovery phase of research, which is often difficult to navigate without a Research Guide.\footnote{Frameworks such as ISEF formalize this: the ``start date'' of a project is defined as when data collection begins, with literature review and study design classified as preparation prior to that start (Society for Science, ISEF Rules FAQ; \url{https://www.societyforscience.org/isef/international-rules/faq/}). We treat research as the complete cycle, including discovery and question formulation.}

The parallel is the Research Software Engineer (RSE). In 2012, a small group at a workshop in Oxford realized that the people who write software for research had no career path, no professional identity, and no name. They coined the term ``Research Software Engineer,'' and within thirteen years built a profession: a name, a manifesto, dedicated funding, an international survey, a professional society, and a growing academic literature \cite{cohen2021}.\footnote{For the movement's own retrospective, see Hettrick, ``A Not-So-Brief History of Research Software Engineers,'' Software Sustainability Institute (2016); \url{https://www.software.ac.uk/blog/2016-08-17-not-so-brief-history-research-software-engineers}.} The first RSE workshop in 2013 attracted 56 people. By 2024, US-RSE alone had thousands of members.

We propose that the Research Guide needs the same infrastructure. This paper is the first step: recognizing and naming the role so that professional infrastructure can follow. The doctoral apprenticeship remains appropriate for its own stage and the paper does not propose to change it; the Research Guide profession is proposed for the stages before and alongside it, where the research process is a learnable practice that no profession currently teaches. Section~\ref{sec:tiers} develops this argument directly.

\begin{center}
\fcolorbox{gray!50}{gray!5}{%
\begin{minipage}{0.94\textwidth}
\small
\textbf{The argument in brief.} We propose the name \textbf{Research Guide} for the person who develops another's capacity to do research. The role is structurally analogous to the school-team coach in sport and the school band director in music: the Tier~2 profession trained in \emph{cognitive apprenticeship} to serve learners with aptitude or serious interest. Music and sport each have a profession for every tier of learner (general, aptitude, pre-professional); research has the classroom science teacher at Tier 1 (covering science content) and the PhD advisor at Tier 3, but no profession teaches the research process itself to the middle tier. The gap is structural, not a failure of effort: the individuals who currently perform this work have no shared credential, training pathway, career structure, or professional identity. Naming the role is the prerequisite for building that infrastructure, as the Research Software Engineer movement demonstrated over the past thirteen years.
\end{minipage}}
\end{center}

% ============================================================================
% 2. The Role That Has No Name
% ============================================================================

\section{The Role That Has No Name}
\label{sec:role}

A Research Guide develops the human capacity for inquiry through a specific set of activities: helping a student formulate a research question that is neither trivially answerable nor impossibly broad; teaching how to search, read, and evaluate existing literature; guiding experimental or analytical design; helping interpret ambiguous results; developing the ability to write clearly about complex findings; and, perhaps most importantly, modeling the dispositions of a researcher: tolerance for ambiguity, persistence through failure, intellectual honesty, and the judgment to know when a result is meaningful and when it is noise. These activities require two distinct skill sets that rarely coexist in the same person.

\subsection{Three tiers of learners and the missing profession}
\label{sec:tiers}

We organize the discussion around three tiers of learners in any craft. This is a rhetorical structure for the professionalization argument, not a novel learner taxonomy: its job is to make visible which tier has a profession behind it and which does not. Each tier is served by a distinct kind of instructor and, in mature domains, by a dedicated profession.

\textbf{Tier 1: general population.} Everyone encounters the domain in classroom form: required general music, required PE, required science. The instructor is a generalist teacher working at classroom scale with a structured curriculum built around content with known answers.

\textbf{Tier 2: students with aptitude or serious interest.} Learners who have chosen to go deeper: the school band, the school sports team, the community league, the serious private lesson. The instructor is a specialist trained in \emph{cognitive apprenticeship}: modeling expert moves on authentic problems whose answers are open, coaching, scaffolding, fading \cite{collinsbrown1991}. The learner does real work; the content is unpredictable, but the teacher's moves for navigating uncertainty are externalized rather than left to observation and osmosis. The instructor does not need to have competed or researched at the elite level, but must be trained in the craft of eliciting real work from others.

\textbf{Tier 3: specific professional focus.} A small number of learners commit to professional-level practice. In music, this is the conservatory track. In sport, the pro academy. In research, the PhD program. The instructor is a master practitioner, and the delivery is pure apprenticeship: intensive, long, often 1-on-1, with tacit transmission at the frontier of the craft. The paper does not propose to change this tier.

The tiers and the pedagogical modes they rely on correlate in practice but are distinct as concepts: the mode describes \emph{technique}, the tier describes \emph{whom} the instruction serves. A Tier 1 classroom does mostly structured instruction; a Tier 2 specialist does mostly cognitive apprenticeship; a Tier 3 master does mostly apprenticeship. The correlation is strong but not rigid.

\textbf{Every mature craft has a profession for every tier.} Sport has a PE teacher for Tier 1, a school/community coach and certified trainer for Tier 2, and an Olympic or pro-academy coach for Tier 3. Music has a classroom music teacher for Tier 1, a school band director or private studio teacher for Tier 2, and a conservatory studio teacher for Tier 3. Research has a classroom science teacher for Tier 1 (teaching science content at classroom scale) and a PhD advisor for Tier 3, but no defined role for Tier 2. Table~\ref{tab:modes-domains} summarizes the gap. The doctoral apprenticeship is the research analogue of elite coaching and the conservatory; the classroom science teacher is the research analogue of the classroom music teacher. What is missing is the research analogue of the school-team coach and the school band director: a profession trained to deliver cognitive apprenticeship for research to the population of learners with aptitude and serious interest. The sport analogy helps illustrate what it takes to find and develop larger numbers of top-level researchers: if you want NBA champions you need a basketball court in every backyard and every school, PE teachers introducing the game to every student, plus tens of thousands of trained coaches who know how to teach the game well to millions.

\begin{table}[t]
\centering
\caption{Three tiers of learners across domains. Each cell names the practitioner who serves that tier. Music and sport each have a profession in every tier; research has professions for Tier 1 and Tier 3 but not for Tier 2. Rows run from broadest reach (Tier 1, top) to narrowest (Tier 3, bottom).}
\label{tab:modes-domains}
\small
\begin{tabular}{@{}p{1.5cm}p{4.1cm}p{4.1cm}p{4.1cm}@{}}
\toprule
& \textbf{Music} & \textbf{Sport} & \textbf{Research} \\
\midrule
\textbf{Tier 1} & Classroom music teacher & PE teacher & Classroom science teacher \\
\arrayrulecolor{gray!40}\midrule\arrayrulecolor{black}
\textbf{Tier 2} & School band/choir director; private studio (Suzuki, MTNA) teacher; community-ensemble director & School-team coach; community/club coach; certified personal trainer & \textit{\textbf{Missing profession}} (proposed: Research Guide) \\
\arrayrulecolor{gray!40}\midrule\arrayrulecolor{black}
\textbf{Tier 3} & Conservatory studio teacher & Olympic coach; pro-academy coach & PhD advisor (research faculty) \\
\bottomrule
\end{tabular}
\par\vskip 4pt
\begin{minipage}{\dimexpr 1.5cm+4.1cm+4.1cm+4.1cm+6\tabcolsep\relax}
\footnotesize \textit{Note.} Tier~1 = general population (structured instruction); Tier~2 = students with aptitude or serious interest (cognitive apprenticeship); Tier~3 = those committed to professional-level practice (pure apprenticeship). Research is the only domain of the three without a Tier~2 profession.
\end{minipage}
\end{table}

Science classes, placed in this schema, are Tier 1 instruction. The canonical classroom lab is a demonstration of known content: the question, method, and expected result are fixed in advance, and the student executes the protocol and compares the obtained result to the expected one. The judgment that defines research is absent because it cannot be graded against an answer key. What the Research Guide profession adds is the Tier 2 layer: scaffolding for a process whose outcomes are, by definition, unknown.

\subsection{The Tier 2 gap in research}
\label{sec:gap}

The tiers describe learners, not institutions. In mature domains, institutions commonly host several tiers simultaneously, with structured pathways between them. A conservatory typically runs a pre-college division (auditioned, mostly Tier 2 with some Tier 3 aspirants) alongside its full bachelor's and master's programs (Tier 3), and a community music division (Tier 2 open enrollment). Specialized performing-arts high schools (Interlochen, Walnut Hill, LaGuardia, Idyllwild) blend Tier 2 and early Tier 3 under the same roof. Sport does the same: IMG Academy and the large tennis and soccer academies host Tier 2 age-group programs alongside Tier 3 pro-development tracks, and a single club or high-school athletic program typically runs teams at multiple competitive levels with athletes moving between them. Students in these systems do not have to commit to Tier 3 to be served well; the profession that staffs the institution supports them at whatever tier they occupy and facilitates movement between tiers as their ambitions clarify. A pre-college violin student with aptitude but uncertain talent can be taught well for years by the same profession that would teach a conservatory-bound peer; the profession serves both.

Research has institutions too. A small set of magnet and specialized secondary schools run research programs of notable quality (Bergen County Academies, Stuyvesant High School, Thomas Jefferson High School for Science and Technology (TJHSST), the BASIS network, the Illinois Mathematics and Science Academy, the North Carolina School of Science and Mathematics, and the Princeton International School of Mathematics and Science (PRISMS)); the historical Soviet fizmat schools and the Nuffield Physics project did the same; a handful of selective summer programs (the Research Science Institute (run by the Center for Excellence in Education, hosted at MIT),\footnote{Center for Excellence in Education, Research Science Institute; \url{https://www.cee.org/programs/research-science-institute}.} the Simons Summer Research Program at Stony Brook, the Secondary School Field Research Program, the Freshman Research Initiative at UT Austin) reach further cohorts. What these institutions lack is a profession. The individuals who staff them are subject teachers, postdocs, or scientists doing research mentoring as a labor of love, without a shared credential, common training, professional organization, or standardized career track. The most successful of these institutions have found individuals who happen to possess both pedagogy and research experience (scientists who became teachers, teachers who pursued research, postdocs who discovered a talent for mentoring), but ``happen to'' is the operative phrase: these practitioners are accidents of biography, not the product of a training program designed for the role, and most of them still work in apprenticeship mode themselves, one mentor with a small group, scale-limited by the instructor-to-student ratios that apprenticeship imposes. The constraint is structural, not a judgment about the work these champions do. When a champion retires, there is no profession from which to hire a replacement, and the program typically contracts or ends. The institutions that exist cannot readily multiply, because the labor market that would allow them to does not exist.

The long-running competition layer makes the same gap visible from a different angle. In the United States, the Science Talent Search (STS, 1942) and the International Science and Engineering Fair (ISEF, 1950) have incentivized pre-college research for more than seventy years. Another competition, the International Young Physicists' Tournament (IYPT), founded in 1988, addresses the same goal with a team-based approach and open-ended problems: teams of secondary-school students spend nearly a year researching seventeen physics questions with no predetermined answers, then defend their solutions against other national teams in structured debate before an international jury \cite{kluiber2007}.\footnote{For IYPT's own description of format and founding date, see \url{https://www.iypt.org/basic-facts/}.} IYPT is especially revealing because its format approximates what Tier 2 research pedagogy should look like (teams, open problems held in common, iterative critique, year-long engagement), and yet IYPT still depends on whichever physics teacher or professor each country happens to find to coach; no profession produces or credentials IYPT coaches any more than it does ISEF mentors. Three long-running competitions, seven decades of documented demand across continents, and still no profession emerged beneath them. The competitions reveal the gap. They do not fill it. Students who reach the top of any of them still do so through a small set of well-resourced magnet schools, ad-hoc university-lab placements arranged through family or school networks, or the selective summer programs noted above: an infrastructure whose access tracks wealth and connection more than talent (Section~\ref{sec:who}).

Sport and music each have a certified Tier 2 profession at every level of the competition ladder, not only at the elite top. US Soccer licenses coaches at graded levels (D through Pro); the American Swimming Coaches Association certifies swim coaches at five levels. In music, the Music Teachers National Association (founded 1876) certifies private studio teachers, state boards certify school music teachers through NAfME alignment, and the Suzuki Association of the Americas certifies method teachers. A parent looking for serious swim coaching or piano instruction for their child knows where to look: the certification exists, the labor market exists, the ladder exists. Research has built the competition ladder (local fair, state fair, STS at the national level, ISEF and IYPT internationally), but the Tier 2 profession that should staff it has no counterpart. The classroom science teacher exists and is state-certified (Tier 1). The PhD advisor exists and is credentialed by doctorate (Tier 3). The coach in between, who would guide a student from the school fair to the state competition to the national and beyond, has no recognized profession, no credential, no career track. A parent looking for serious research mentoring has no equivalent place to turn.

The sport and music parallels are not a perfect symmetry. Sport develops physical craft; music develops performance craft; research develops intellectual craft, and the three are not interchangeable. The age at which Tier 3 training typically begins differs sharply: in music, pre-professional apprenticeship can start at four or six (Suzuki, conservatory pre-college programs); in sport, often at ten or twelve (club teams, pro academies); in research, almost always at twenty-two or later (the PhD program). This age asymmetry is itself informative. Music and sport have Tier 2 professions that identify aptitude early and feed Tier 3 systematically. Research does not, so Tier 3 waits until the student finds their own way to a PhD program, usually by chance of parental or school resources. The late start is not intrinsic to the craft. Many phenomena admit authentic research at any age: local ecology, behavioral observation, community-level problems, environmental variation, computational experiments on open datasets (Section~\ref{sec:modes} discusses which domains are accessible at which population). The Soviet fizmat schools and many top ISEF projects demonstrate that teenage learners can do authentic research at high levels when domain and method are chosen well. The late start is a consequence of the missing Tier 2 profession, not of the craft itself.

\subsection{The pedagogy--methodology skill-set gap}
\label{sec:pedmethod}

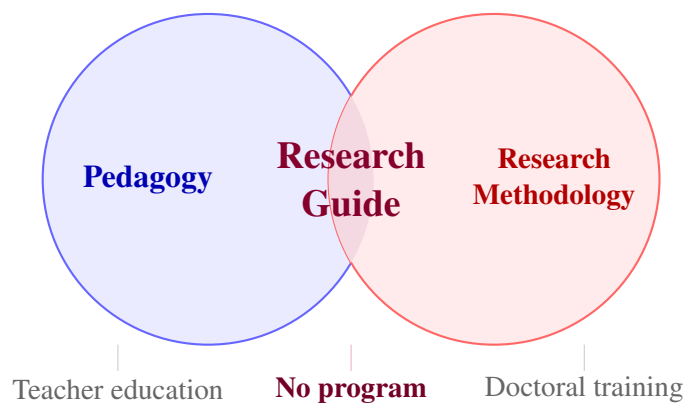
\begin{figure}[t]
\centering
\begin{tikzpicture}[font=\small]
  \def\sep{1.9}   % center-to-center half-distance
  \def\rad{2.2}   % circle radius

  % Left circle - Pedagogy
  \fill[blue!8] (-\sep,0) circle (\rad cm);
  \draw[blue!60, thick] (-\sep,0) circle (\rad cm);
  \node[align=center, font=\large\bfseries, blue!70!black] at (-2.7,0) {Pedagogy};

  % Right circle - Research Methodology
  \fill[red!8] (\sep,0) circle (\rad cm);
  \draw[red!60, thick] (\sep,0) circle (\rad cm);
  \node[align=center, font=\normalsize\bfseries, red!70!black] at (2.7,0) {Research\\Methodology};

  % Intersection fill
  \begin{scope}
    \clip (-\sep,0) circle (\rad cm);
    \fill[purple!15] (\sep,0) circle (\rad cm);
  \end{scope}

  % Intersection label
  \node[align=center, font=\Large\bfseries, purple!70!black] at (0,0) {Research\\Guide};

  % Bottom labels — well below the circles, larger font
  \node[font=\normalsize, gray!80!black] at (-3.1,-2.8) {Teacher education};
  \node[font=\normalsize\bfseries, purple!60!black] at (0,-2.8) {No program};
  \node[font=\normalsize, gray!80!black] at (3.1,-2.8) {Doctoral training};

  % Thin lines connecting labels to circles
  \draw[gray!40, thin] (-3.1,-2.55) -- (-3.1,-2.2);
  \draw[purple!30, thin] (0,-2.55) -- (0,-2.2);
  \draw[gray!40, thin] (3.1,-2.55) -- (3.1,-2.2);
\end{tikzpicture}
\caption{The pedagogy--methodology gap. No existing training program produces both skill sets together. The Research Guide profession sits at the intersection; it is the Tier~2 profession identified as missing in Table~\ref{tab:modes-domains}. What the Research Guide needs on the right-hand side is research \emph{fluency}, not the specialty depth of a doctoral specialist; doctoral training is named below only as the current source of substantial research experience, not as the target depth.}
\label{fig:gap}
\end{figure}

The gap in institutions and in the competition ladder points to a deeper gap in the hiring pool. The Tier 2 role requires two skill sets that rarely coexist in the same person (Figure~\ref{fig:gap}).

The first is \textbf{pedagogy}: how to teach, scaffold learning, assess understanding, adapt to different learners, and motivate students through difficulty. This is the domain of teacher education.

The second is \textbf{research methodology}: how to formulate questions, design experiments, collect and analyze data, navigate peer review, and situate findings within existing knowledge. A Research Guide needs \emph{research fluency} (enough firsthand research experience to guide students through the cycle on open problems), not the narrow-specialty depth of a PhD in a particular subfield.

No existing training program produces both. Science teachers ``rarely have an opportunity to participate in scientific research'' \cite{aaas2020}, and although the benefits of research experience for teachers are documented, ``few programs extend this educational vehicle to preservice or in-service teachers, and even fewer have studied the impact'' \cite{krim2019}. The National Science Foundation funds Research Experiences for Teachers (RET), but these are small-scale summer supplements, not a standard component of teacher preparation. On the other side of the gap, researchers and faculty are trained in methodology but not in mentoring: in a randomized controlled trial of research mentoring, only 21\% of mentors reported any prior mentoring training \cite{pfund2014}, and the NASEM report on effective mentorship confirmed the broader pattern that most faculty who mentor research students have never been trained to do so \cite{nasem2019}.

The structural consequence is stark. When a school or university needs someone to teach students how to do research, the hiring pool consists of two populations, neither of which is prepared for the role:
\begin{itemize}[leftmargin=*]
    \item A teacher who knows how to teach but has never done research.
    \item A researcher who has done research but has never been trained to teach.
\end{itemize}

\subsection{Why the status quo fails}
\label{sec:statusquo}

Because there is no profession to hire from, research programs are structurally fragile. When the person who runs a school's research program leaves, the program is at risk of collapse, not because the institution lacks commitment, but because there is no labor market from which to recruit a replacement. With the exception of the best-resourced magnet and specialized schools, most schools that offer research have a single person doing it; the availability of qualified teachers is widely recognized as the bottleneck for expanding authentic science research programs. In the parallel context of graduate education, when a principal investigator changes institutions, retires, or leaves academia, entire research groups are at risk of collapse.

The fragility is compounded by funding. Research mentoring programs rarely appear as permanent budget lines; they are sustained through grants, discretionary funds, or the uncompensated labor of individual champions. When budgets tighten, programs without a professional category are the first to lose funding: there is no line item to defend, no professional standard to invoke, and no constituency organized to advocate. A math department does not need to justify why it employs math teachers; a research program built on soft money must rejustify its existence every budget cycle.

Course-Based Undergraduate Research Experiences (CUREs) face the same personnel problem from a different angle. Institutional opposition is not the main barrier: 68\% of CURE faculty report that teaching a CURE contributed positively toward their tenure and promotion \cite{jmbe2020}. The limiting factor is the availability of faculty willing and able to do the work: CUREs demand time, cost, and pedagogical effort that standard teaching loads do not accommodate \cite{jmbe2020}.

The NSF's response to mentoring quality problems in REU programs is telling: NSF now requires that ``REU grantees must ensure that research mentors receive appropriate training or instruction'' \cite{cbe2024}, implicitly acknowledging that the mentors are not trained for the role they are being asked to perform. The requirement is a patch, not a solution: it asks individual programs to compensate for the absence of a profession. No funding agency requires that math teachers receive ``appropriate training or instruction'' in mathematics, because the profession of math teacher already exists.

A natural question is whether research mentoring is already part of the professoriate. The professoriate is a Tier 3 profession: its members were trained through doctoral apprenticeship and are credentialed to reproduce that mode with their own PhD students. When the same practitioners mentor at Tier 2 (middle and high school students, undergraduates, or adults learning to do research outside the doctoral track), they default to the mode they experienced as students: pure apprenticeship at its one-mentor-to-a-few ratio, with methodology transmitted through extended personal immersion. No training program prepares faculty for Tier 2, which calls for a different pedagogy at a different scale (cognitive apprenticeship, Section~\ref{sec:tiers}). The result is that a Tier 2 function is often performed by Tier 3 practitioners using Tier 3 methods, and the structural mismatch limits reach and consistency.

The parallel question is whether research mentoring is already part of classroom science teaching. Classroom science teaching is a Tier 1 profession: its members are trained and credentialed to deliver science content with known answers to a full classroom through structured instruction. When classroom teachers are asked to guide students through authentic research at Tier 2 (for example, in AP Research or the IB Extended Essay), they default to what they were trained in: content delivery, structured problem sets, and assessment against predetermined rubrics, on top of a full teaching load. No training program prepares them for Tier 2 either. The result, symmetric to the Tier 3 case, is a Tier 2 function performed by Tier 1 practitioners using Tier 1 methods: the framework is real, the teachers' effort is real, but neither produces authentic research at the scale the craft calls for.

The proposal is not to replace faculty mentoring or teacher-led research but to recognize that Tier 2 work is a distinct specialization requiring its own training, its own career structure, and its own professional identity. As long as the Tier 2 role remains unnamed and unrecognized, the problems described in this section will persist.

\subsection{Why Tier 2 assessment requires a profession}
\label{sec:assessment}

The structural mismatch shows up most sharply in how the three domains evaluate a developing learner. Music is the closest parallel to research because music assessment, like research assessment, rests on a trained professional's judgment of qualities that cannot be reduced to an objective measurement. The Associated Board of the Royal Schools of Music (ABRSM) graded music examinations, organized into Grades 1--8 in their current form since 1933, are administered by examiners who listen to each student perform and mark not only accuracy and fluency but the qualities that make a performance musical. The Distinction-grade descriptors list ``well projected tone, sensitive use of tonal qualities, expressive idiomatic musical shaping and detail, assured and fully committed performance, vivid communication of character and style.''\footnote{ABRSM marking criteria for graded music exams; \url{https://us.abrsm.org/en/exam-support/your-guide-to-abrsm-exams/how-we-mark-exams/}.} The ABRSM grade is the notation of the examiner's professional judgment about those qualities, not a score generated by a stopwatch or an algorithm. The studio teacher preparing the student works on the same qualities week after week.

Sport, by contrast, is a weaker parallel for the grading ladder because its measurable credentials (swim times, track records) can exist without a professional's judgment in the way music's cannot. It remains a strong parallel for one deeply qualitative Tier 2 function: talent identification. A coach recognizes which swimmers, players, or gymnasts have the potential to move toward Tier 3 on signals (coachability, rate of improvement, decision-making under pressure, athletic disposition) that no stopwatch and no rubric can capture. That recognition is precisely the Tier 2 work, and it is exactly what research currently has no profession to perform: the classroom science teacher is not trained to spot who might become a researcher, and the PhD advisor engages students only after they have already arrived at graduate school.

Research has the Tier 1 rubric and the Tier 3 peer-reviewed publication, and essentially nothing in between. Course-level instruments like AP Research \cite{collegeboard2026} and the IB Extended Essay \cite{ibo2024} grade a finished artifact once a year against a generic rubric, not a developmental record of how a student's research capability grew. At the college level, the American Association of Colleges and Universities (AAC\&U) \emph{Inquiry and Analysis} VALUE rubric is the most widely adopted instrument of its kind and, in its own framing language, ``addresses the \emph{products} of analysis and inquiry, not the \emph{processes} themselves'' and is ``intended for institutional-level use in evaluating and discussing student learning, not for grading'' \cite{aacu2018}. Tier 3 peer review is summative rather than developmental assessment: binary, artifact-based, and arriving years after the work a developing researcher would most benefit from having read. Other instruments have been built to fill the middle, including research-skill-development frameworks and undergraduate-research self-report surveys \cite{willison2018,lopatto2004,weston2015}. Each was carefully designed; each captures a real slice of the problem. What none of them has is a profession behind it to wield the instrument continuously, calibrate it across instructors, and update it across cohorts. Music built ABRSM grades after it had conservatory and studio teachers; research cannot build the analogue until it has Research Guides.

Three propositions make the structure of this argument explicit. First, Tier 2 tasks are non-standardizable by design: open problems have unknown answers, and a rubric can only cover what can be pre-specified, so any rubric broad enough to span authentic research becomes generic and any rubric tight enough to be informative becomes project-bound and non-reusable. Second, what closes the gap is the trained professional's judgment applied repeatedly: the studio teacher reading a student's weekly progress across a year, the coach watching whether a swimmer's stroke is improving, the research mentor watching whether a student's second question improves on the first. The formal grade or verdict is the \emph{notation} of that judgment, not a substitute for it; the same structure holds at Tier 3, where peer review works because the reviewer's judgment is the assessment and accept/reject is its notation. Third, without a profession, individual judgments do not converge: the profession creates the baseline through shared training, shared vocabulary, and shared expectations calibrated across practitioners. A rubric without a profession is a document; inside a profession it is a vocabulary.

Metrics in music and sport can be gamed. Studios sometimes drill students to pass grades without building musicality, and athletes can train narrowly to hit recorded metrics without building broader athletic capacity. Calibrated professional judgment is structurally more resistant. It is multidimensional (optimizing one quality leaves another exposed), calibrated across practitioners (deceiving one examiner is plausible; deceiving a profession taught to recognize tells is not), longitudinal (gaming shows up as discontinuity over months or years), and self-updating (the profession incorporates newly observed gaming patterns into what it watches for). Goodhart's Law (when a measure becomes a target, it ceases to be a good measure) applies strongly to metrics and much more weakly to professional judgment.

A continuum of tiers also enables a continuum of credentials. Music carries two parallel ladders: one for learners (ABRSM Grades 1--8) and one for the professionals who teach them (Music Teachers National Association certifications, Suzuki Association credentials, state music-education licensure). Each ladder is continuous because the profession that sustains it is continuous. Research has neither: no intermediate credential for a student between a high-school research project grade and a journal author line, and no professional credential at Tier 2 at all. The RG-1/RG-2/RG-3 career track proposed in Section~\ref{sec:profession} would supply the professional ladder. A learner ladder is harder to sketch; related work by the first author on opening the scientific process points to where its evidence base would come from: a visible, trackable record of what a student has actually done (questions framed, methods designed, analyses replicated, errors corrected) is richer developmental evidence than a publication count or a single holistic score, in the same way that a studio teacher's running log of repertoire and technique is richer evidence than one end-of-year recital \cite{samsonau2026open}.

\subsection{Evidence that Tier 2 work scales and benefits learners}
\label{sec:evidence}

Can cognitive apprenticeship for research reach Tier 2 at all, or must it remain a rare hand-craft? It can. The definition of cognitive apprenticeship \cite{collinsbrown1991} does not fix the instructor-to-student ratio; because the expert's processes are externalized rather than transmitted by osmosis, the mode admits higher ratios than pure apprenticeship does. Working programs already demonstrate this. The Freshman Research Initiative (FRI) at UT Austin has engaged thousands of first-year undergraduates in course-based research across roughly 25 research streams, each led by a PhD-level research educator as a professional role; a propensity-matched evaluation found measurable gains in STEM graduation and on-time completion for students who completed the full three-course sequence, with effects similar across demographic groups \cite{rodenbusch2016}. Smaller programs show complementary evidence in different domains, including the AI for Scientific Research program at NYU, which the first author founded, running teams of college students on real AI-for-science problems with external research laboratories using a structured role architecture and staged project lifecycle \cite{samsonau2024aifsr}.

The benefits accrue to the full Tier 2 population, not only to those who continue to Tier 3. Beyond the STEM-completion gains documented at FRI \cite{rodenbusch2016}, studies also report gains in scientific literacy \cite{gormally2009} and psychological well-being through autonomy, competence-building, and connection to others who share their interest \cite{walkington2022}. Evidence extends down to middle school, where science-fair participation has been shown to improve student understanding of the scientific-inquiry process and attitudes toward STEM careers \cite{schmidt2017}. In the age of AI (Section~\ref{sec:ai}), these are precisely the capacities that remain most valuable to develop.

% ============================================================================
% 3. Who Needs Research Guides
% ============================================================================

\section{Who Needs Research Guides}
\label{sec:who}

\subsection{Middle and high school students}

Approximately 15 million students are enrolled in U.S. high schools, and demand for research experience among them is both substantial and growing: AP Research enrollment grew 15-fold between 2016 and 2025, from roughly 2{,}800 to over 43{,}000 exam takers \cite{collegeboard2026}; the International Science and Engineering Fair draws approximately 175{,}000 students through 400 affiliated fairs each year;\footnote{Society for Science press release, 2024; \url{https://www.societyforscience.org/press-release/columbus-to-host-regeneron-isef-2025/}.} and the IB Extended Essay reaches nearly 100{,}000 students worldwide annually \cite{ibo2024}. Of the high school students who encounter research at all, most encounter it only through structured frameworks (such as AP Research and the IB Extended Essay), which do not require teachers to have research experience, do not provide external mentorship, and operate within a single academic year.

Schools that prepare students seriously for authentic research beyond those frameworks are far rarer. In the 2020--2026 Science Talent Search, only five schools had a finalist every second year, clearly differing from the rest of approximately 27{,}000 U.S. high schools in how they approach research education.\footnote{Authors' analysis of 280 STS Top 40 finalists from the 2020--2026 cohorts; underlying data at \url{https://www.societyforscience.org/regeneron-sts/}.} These five are research-focused magnet and independent secondary schools with dedicated research programs and research-experienced staff. Their advantage is institutional position: substantial resources, dedicated research facilities, and established ties to local universities and research organizations. Such schools already operate something close to a Research Guide role, but typically in the pure-apprenticeship mode of Tier 3 (Section~\ref{sec:tiers}), constrained to the one-mentor-to-a-few ratios (Section~\ref{sec:statusquo}) that limit scale. The same resource-and-network combination constrains access at the individual level: participation in science fairs depends on ``local resources, teacher time, mentorship, lab space and travel funding,'' and ``participants and winners who advanced to state and national competitions consistently came from a small number of well-resourced schools'' \cite{pmc2021}; even entering a science fair, Zimmer observed, requires ``the flexibility to spend hours on paperwork and a social network of scientists'' \cite{zimmer2016}.

The problem is not that students at high schools other than those five lack interest in research or are less talented. The problem is that the system has no mechanism to connect interested students with qualified mentors, because the profession of ``qualified research mentor'' does not exist. No study has measured how many students would pursue research given the opportunity and resources. We know how many play varsity sports, take AP exams, or study music, because those fields have professional infrastructure that tracks participation. Research mentoring has no such infrastructure.

\subsection{College students}

Only about 22\% of graduating seniors self-report having worked with a faculty member on a research project, with rates varying from roughly 16\% to 43\% across institution types \cite{nsse2019}, though NSSE's question was broad enough to include classroom projects as well as original contributions to the field \cite{schneider2021}. Institutional records at individual universities, using narrower definitions based on course enrollment, yield lower annual rates: about 5\% at the University of Georgia \cite{fechheimer2011} and 10--14\% at UC Berkeley \cite{berkes2008}. The access problem is familiar: faculty prefer students with prior research experience, but students cannot acquire research experience without faculty willing to mentor beginners.

Federal funding for the primary pipeline, NSF's Research Experiences for Undergraduates (REU), has faced repeated cuts. REU supports several thousand students per year, a fraction of the millions enrolled in undergraduate STEM programs.\footnote{U.S. National Science Foundation, Research Experiences for Undergraduates program; \url{https://www.nsf.gov/funding/pgm_summ.jsp?pims_id=5517}.} Community college students, who represent nearly half of U.S. undergraduates, are largely excluded: their institutions have minimal research infrastructure, their faculty carry heavy teaching loads, and their students often work full-time.

\subsection{Adults outside academia}

Beyond formal education, millions of adults want to do research, not as a career, but as a serious intellectual pursuit. The demand is not hypothetical. Citizen science platforms such as iNaturalist, eBird, and Zooniverse have collectively engaged millions of participants in authentic scientific data collection \cite{nasem2018cs}. The appetite for participation in real science is enormous and growing.

Yet the landscape reveals a structural gap. It ranges from citizen science (large-scale data collection for professional researchers' studies) through participatory science (co-designed fieldwork and deeper collaboration) to fully independent investigation. Citizen science participants are disproportionately well-educated \cite{nasem2018cs}, but the vast majority remain data collectors. The gap between contributing observations to someone else's study and designing your own investigation is precisely the gap a Research Guide would bridge.

% ============================================================================
% 4. What Research Guides Do: Stages and Skills
% ============================================================================

\section{What Research Guides Do}
\label{sec:stages}

The Research Guide is not a single role performed the same way at every level. The core process is the same (guiding people through authentic research as a complete process, from motivation through methodology to communication) but the emphasis, the accessible domains, the depth of claims, and the mode of inquiry change across populations and projects.

One element, however, is constant: \emph{curiosity}. Curiosity is not a phase that learners outgrow. Neuroscience research has shown that states of curiosity modulate hippocampus-dependent learning via the dopaminergic circuit: curious learners form stronger memories and make deeper connections, regardless of age \cite{gruber2014,kidd2015}. The Research Guide's first responsibility is to identify, protect, and channel curiosity into systematic inquiry. In practice this means helping the student arrive at a question that is recognizably their own: ownership of the question is what keeps the engine running once the work gets hard. The populations below differ in how the \emph{form} of guidance changes; curiosity is the engine that drives all of them.

\subsection{Research across populations}

\begin{figure}[t]
\centering
\begin{tikzpicture}[
    stage/.style={draw, rounded corners=4pt, thick,
                  minimum width=2.8cm, minimum height=1.0cm,
                  align=center, font=\small\sffamily, inner sep=2pt},
    highlight/.style={fill=purple!15, draw=purple!60, very thick,
                      minimum width=4.6cm, minimum height=1.5cm,
                      font=\normalsize\sffamily\bfseries},
    plain/.style={fill=blue!8, draw=blue!30},
    arrow/.style={-{Stealth[length=5pt, width=4pt]}, thick, gray!60}
]
  \def\R{2.8}
  \def\Rq{3.6}

  % Discovery \& Question Formulation is placed at a larger radius to avoid crowding.
  \node[stage, highlight] (Q)  at ( 90:\Rq) {Discovery \&\\[-1pt] Question Formulation};
  \node[stage, plain]     (M)  at ( 30:\R) {Method};
  \node[stage, plain]     (D)  at (-30:\R) {Data};
  \node[stage, plain]     (A)  at (-90:\R) {Analysis};
  \node[stage, plain]     (C)  at (-150:\R){Conclusion};
  \node[stage, plain]     (Co) at ( 150:\R){Communication};

  % Clockwise arcs between node centers — uniform bend = rotational symmetry.
  \foreach \from/\to in {Q/M, M/D, D/A, A/C, C/Co, Co/Q} {
    \draw[arrow] (\from) to[bend left=22] (\to);
  }
\end{tikzpicture}
\caption{The general research cycle shared across modes of inquiry. Every mode (Section~\ref{sec:modes}) specializes these stages: in the hypothetico-deductive mode, \emph{Discovery \& Question Formulation} becomes hypothesis formulation and \emph{Method} becomes the deduction of a testable prediction and its controlled test; in exploratory work, \emph{Discovery \& Question Formulation} is systematic observation without a prior hypothesis; in tool development, it is identifying a capability gap. What is universal is that research is a complete process, not a single step. Across all modes, discovery and question formulation (highlighted, enlarged) can easily consume half or more of the total project time, yet it is the stage least addressed by standardized frameworks.}
\label{fig:cycle}
\end{figure}
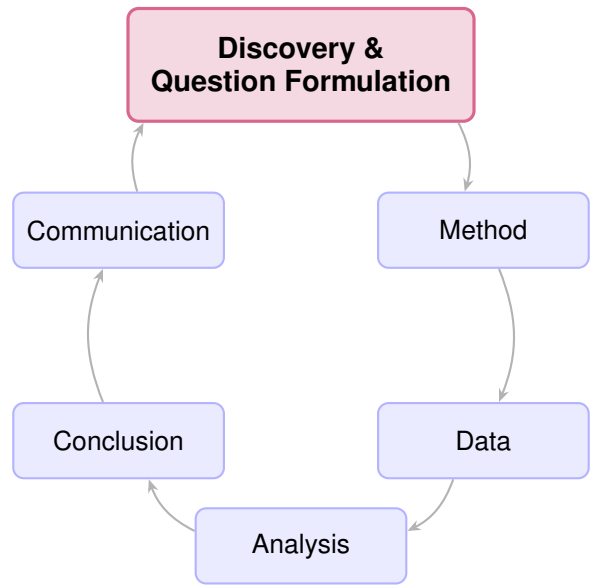

\begin{table}[!htbp]
\centering
\caption[Research across populations]{Populations served by the Research Guide profession (Tier~2 in Table~\ref{tab:modes-domains}): same research cycle, different accessible domains and equipment, depth of claims, and Research Guide emphasis. Graduate (PhD) training is Tier~3 and is not listed here; it is served by the existing advisor-apprenticeship profession.}
\label{tab:populations}
\small
\begin{tabular}{p{2.2cm}p{3.5cm}p{2.5cm}p{4.3cm}}
\toprule
\textbf{Population} & \textbf{Accessible domains and typical equipment} & \textbf{Novelty \& depth} & \textbf{Research Guide emphasis} \\
\midrule
\textbf{Middle school} & Ecology, environmental observation, behavioral observation, game design, local community problems. Typical equipment: household materials, smartphones, hobbyist microscopes, field sites. & Original questions; descriptive and early explanatory claims. & Protects curiosity. Full cycle with accessible methods. Teaches that ``I wonder why\ldots'' is the beginning of research, not a homework assignment. \\[2pt]
\arrayrulecolor{gray!40}\midrule\arrayrulecolor{black}
\addlinespace[2pt]
\textbf{High school} & Adds biology, chemistry, physics, data science; computational projects become feasible. Typical equipment: consumer-grade sensors, open datasets, laptops with open-source libraries, modest school labs. & Original questions and novel applications of known methods; explanatory claims with simple controls and error analysis. & Deepens methodological rigor: controls, literature, statistical reasoning, scientific writing. Models researcher dispositions. \\[2pt]
\arrayrulecolor{gray!40}\midrule\arrayrulecolor{black}
\addlinespace[2pt]
\textbf{Undergraduate} & Adds institutional equipment, specialized instrumentation, IRB-governed human subjects research, cross-lab collaborations, course-based research (CUREs). & Novel contributions at a scope appropriate to coursework or an independent study. & Develops foundational research fluency (question formulation, methodology, evidence evaluation, communication) and scaffolds the shift toward independent judgment. \\[2pt]
\arrayrulecolor{gray!40}\midrule\arrayrulecolor{black}
\addlinespace[2pt]
\textbf{Adults outside academia} \newline {\scriptsize (domain experts, career changers, citizen and participatory scientists)} & Domain expertise opens specialized questions. Home and field instrumentation, open datasets, and participatory-science programs enable authentic work without an institutional lab. & From contributing observations to designing and conducting one's own investigation. & Bridges the gap between data collection and independent inquiry. Provides methodological frameworks for people with domain expertise but no research training. \\
\bottomrule
\end{tabular}
\par\vskip 6pt
\begin{minipage}{\dimexpr 2.2cm+3.5cm+2.5cm+4.3cm+8\tabcolsep\relax}
\footnotesize \textit{Note.} Research is a complete process, and multiple modes of inquiry (Section~\ref{sec:modes}) are possible for all four populations; only the instrumentation and depth of claims vary. The domains and equipment listed above are illustrative, not prescriptive. At the middle and high school levels, these domains are accessible with household materials and consumer-grade equipment at costs comparable to other structured extracurriculars; a research-grade laboratory is not required. These are not stages on a ladder: a middle schooler studying local ecology is doing real research, not a simplified version of what a graduate student does.
\end{minipage}
\end{table}

The critical insight is that the research is real for every population. Not simplified, not imitation, not a classroom exercise dressed up as inquiry. As Dewey argued, inquiry is a natural human capacity, not a specialized activity reserved for experts \cite{dewey1910}. A middle schooler studying whether local pond water quality varies by season is doing research as real as a graduate student investigating protein folding. Both follow a complete process from motivation through methodology to communication; the specific cycle depends on the mode of inquiry (see Section~\ref{sec:modes}). Across all modes, question formulation may be the most important phase: discovering what is known and what is not known, and framing a question worth pursuing. Einstein and Infeld, in their 1938 overview of physics, emphasized that formulating a problem well is often more consequential than solving it. Formulation is also where student engagement is won or lost: a project a student has framed themselves carries an ownership that an assigned question cannot replicate, and this autonomy is a primary driver of the sustained motivation on which authentic research depends \cite{walkington2022}. The discovery and question formulation phase often consumes most of the time in authentic research, yet standardized frameworks like AP Research and the IB Extended Essay focus primarily on what comes after: methodology, data collection, and analysis. This inversion is precisely what distinguishes the Research Guide's work from conventional instruction. What changes across populations is the domain knowledge required to ask certain questions, the depth of causal claims the researcher can credibly make, and the instrumentation available. What changes across modes is the cycle itself: exploratory research begins with observation, not a hypothesis; tool development begins with a capability gap, not a question about nature; data-driven discovery begins with a dataset, not a theory.

Concrete examples clarify what authentic research at these tiers can look like without a research-grade laboratory. A middle-schooler with a plate of water, a shaker of pepper, and several kinds of soap can investigate whether surfactant patterns are programmable: does the order, spacing, or type of soap droplet predict the resulting two-dimensional pattern, or does the system become chaotic when more than one source is present? A high-schooler with a laser pointer, a heat source, and a smartphone camera can tackle an inverse problem: given only the dancing speckle pattern projected on a wall, what can be recovered about the intervening thermal disturbance, and how close can that come to modeling the atmospheric optics that limits every ground-based telescope? An undergraduate with a laptop and open-source machine-learning libraries can record smartphone video of soap films and train a small network on the early-phase interference dynamics to test whether the first seconds of color shifts predict when each specific bubble will collapse. None of these requires a research-grade laboratory or institutional access: the materials are household items, consumer-grade electronics, and open-source software libraries, at a total cost comparable to or less than what families routinely spend on music lessons, competitive sports, or other structured extracurricular activities. Each targets a question whose answer is genuinely unknown; each supports the full research cycle (Figure~\ref{fig:cycle}) from observation through communication. Catalogs of open problems at this tier include the International Young Physicists' Tournament's annual problem set;\footnote{IYPT annual problem set; \url{https://www.iypt.org/basic-facts/}.} specialized research-mentoring organizations also curate accessible projects in this space.\footnote{See, for instance, \url{https://teenscientists.org/}.}

Participation in authentic research also extends beyond the cycle itself: science writing, science-inspired art, and research project management are all forms of genuine engagement with the research process. The Research Guide profession would redefine ``popular science'' from communicating science \emph{to} the public to the public \emph{doing} science. This has implications beyond education: emerging evidence suggests that public engagement with science is more effectively cultivated through participation than through one-way communication \cite{vitone2016}.

\subsection{Modes of research}
\label{sec:modes}

A deeper problem underlies the current landscape: research education overwhelmingly imposes a single model, the hypothetico-deductive cycle (observation/question $\to$ hypothesis $\to$ deduced prediction $\to$ controlled test $\to$ evaluation), regardless of population, age, or context, and in practice a truncated form of it. Education typically collapses the cycle to hypothesis $\to$ experiment $\to$ result, under-emphasizing the problem-finding and question formulation that should precede the hypothesis, and treating the hypothesis as the driver rather than the artifact of careful observation. The collapse has a second cost beyond methodology: removing question formulation also removes the student's opportunity to own a question of their own, and with it one of the strongest drivers of engagement in the work. The dominant approach does not differentiate the \emph{kind} of research; it only simplifies the \emph{difficulty}. Cookbook labs prescribe the same procedure for millions of students, producing a performance of methodology rather than its practice. ``Research methods'' courses are frequently statistics courses in disguise, and assessment rubrics apply the same criteria regardless of domain or type of inquiry.

But science has never been one kind of activity. Historians and philosophers of science (most influentially Crombie and Hacking) have long argued that the European scientific tradition comprises several distinct styles of reasoning, each with its own types of objects, kinds of evidence, and standards of explanation. Empirical work corroborates this heterogeneity: a substantial fraction of findings in large corpora of funded biomedical research fall outside the original grant proposal's scientific categories \cite{aslan2024}, and the literature on scientific serendipity documents that unexpected discovery takes several structurally distinct forms \cite{yaqub2018}. For the Research Guide, at least five modes are practically significant.

\textbf{Exploratory and descriptive research} asks: \emph{What is happening here?} Systematic observation, documentation, and pattern identification without a prior hypothesis. Darwin's observations on the \emph{Beagle}, Goodall's studies of chimpanzee behavior, and the natural history tradition that preceded and enabled modern biology are all exploratory work. It is the most natural entry point for curiosity (look carefully, document what you see, find the pattern), and the mode most systematically excluded by the hypothesis-first requirement that dominates science education.

\textbf{Hypothetico-deductive investigation} asks: \emph{Does X cause Y?} The researcher formulates a hypothesis, derives a testable prediction by deduction, designs a controlled test, and evaluates the outcome against the prediction, corroborating or falsifying it. It is the mode that education already teaches, and often the only one it teaches. It is powerful and indispensable, but it is not all of science.

\textbf{Theory construction} asks: \emph{What would explain this pattern?} Building explanatory frameworks that unify disparate observations. Darwin's theory of natural selection, plate tectonics, and germ theory were each theoretical contributions distinct from the empirical observations they explained. Theory construction requires its own methodology, and multiple disciplines have diagnosed themselves as ``theory-poor'' \cite{muthukrishna2019}.

\textbf{Tool and method development} asks: \emph{What would make this research possible?} The design of instruments, software, protocols, and methods that enable new research. The microscope did not answer a question about cells; it made questions about cells askable. PCR did not discover a gene; it made every gene discoverable. Galison showed that instrument builders constitute a distinct scientific subculture with their own training and modes of demonstration \cite{galison1997}. Dedicated venues (\emph{Nature Methods}, \emph{Journal of Open Source Software}, \emph{HardwareX}) reflect the growing recognition that tool and method development is a first-class contribution, yet education treats tools as given infrastructure rather than designable contributions.

\textbf{Computational modeling and simulation} asks: \emph{What does the model predict?} Building mathematical or computational representations of systems too complex for analytical solution: climate models, epidemiological simulations, agent-based models in ecology and economics. The researcher's contribution is the model itself (its assumptions, its structure, its validation). What distinguishes modeling from mere calculation is that the modeler encodes understood mechanisms into the model: the SIR model encodes how infection spreads, the Lotka-Volterra equations encode predation dynamics, climate models encode radiation physics. The model is transparent: one can ask not only \emph{what} it predicts but \emph{why}.

Artificial intelligence and machine learning do not constitute a separate mode of research but amplify every existing mode: pattern detection at computational scale for exploratory work, tool development (AlphaFold \cite{jumper2021}), theory construction through symbolic regression \cite{lemos2023}, and hybrid physics-ML models. The most philosophically significant development is opaque prediction: deep learning systems that predict outcomes, such as plasma disruptions in fusion reactors \cite{katesharbeck2019}, without modeling the underlying physics. Whether such prediction-without-explanation constitutes scientific understanding is actively debated \cite{vanfraassen1980,boge2022}. For the Research Guide this debate is practical: in every project where a student reaches for machine learning, is the black-box prediction sufficient, or must the student seek to understand \emph{why}? Helping students navigate that judgment is a new competency the profession requires.

These modes are not a hierarchy or stages of a single method; they are parallel contributions. The Research Guide must recognize which fits each student's question and interest. A student who naturally builds things should not be forced through hypothesis testing when her thinking leads toward tool development; a student who observes patiently and notices patterns should not be told her work is not research because she lacks a hypothesis; a student drawn to computational thinking should not be shoehorned into a wet lab. By teaching only one mode, the current system systematically excludes students whose scientific thinking takes other forms, and in doing so narrows the range of contributions that science receives.

The diversity of populations who need research guidance (middle schoolers exploring ecology, high schoolers designing computational experiments, undergraduates accessing institutional equipment, a retired engineer investigating corrosion patterns from her career, citizen scientists transitioning from data collection to hypothesis-driven inquiry) is itself evidence that this is a profession, not a skill. Each requires a different Research Guide emphasis, but the same professional competencies, the way a physical therapist works differently with a child athlete, a post-surgery patient, and an elderly person but remains the same profession.

We do not propose that every student needs a Research Guide, any more than every child needs a conservatory-level music instructor or a professional athletics coach. But the profession must exist for those who do. The question is whether the people who \emph{want} to do research can find a professional trained to guide them, or whether they must rely on whatever ad hoc support happens to be available. Today, the answer is overwhelmingly the latter.

\subsection{The skill set of a Research Guide}

The role requires a specific combination of competencies that no existing training program develops together:

\begin{itemize}[leftmargin=*]
    \item \textbf{Research methodology and modal fluency.} The Research Guide must have conducted research, not merely read about it. Much of research expertise is tacit: what it feels like to have an experiment fail, to stare at confusing data, to realize your question was wrong. This cannot be transmitted from a textbook. It must be experienced firsthand and then communicated to others through guided practice. Beyond this, the Research Guide must be fluent across the modes of research described in Section~\ref{sec:modes}: able to recognize whether a student's question calls for exploratory observation, hypothesis testing, theory construction, tool development, or computational modeling, and to adjust guidance accordingly. A teacher is trained in differentiated instruction, adapting pedagogy to how each student learns. A Research Guide must differentiate the \emph{form of inquiry} itself, matching the mode of research to the student's question and interest. This is a diagnostic skill: a student who says ``I want to study coral reefs'' might need to observe and document (exploratory), test whether a specific pollutant affects coral growth (hypothesis-driven), build a low-cost underwater sensor (tool development), or construct a simulation of reef ecosystem dynamics (computational modeling). The Research Guide's first task is to help the student discover which mode fits, not to funnel every question through a single methodology.

    \item \textbf{Pedagogy.} Knowing how to do research is not the same as knowing how to teach research. The Research Guide must be able to scaffold learning, assess where a student is in their development, and calibrate challenge to the student's zone of proximal development \cite{vygotsky1978}.

    \item \textbf{Developmental assessment.} Unlike a course instructor who grades assignments, a Research Guide assesses the development of complex, ill-structured capabilities (question quality, methodological reasoning, critical reading, persistence) over time. No standardized assessment exists for this. The profession needs one.

    \item \textbf{Risk and productive struggle management.} A Research Guide actively manages two kinds of difficulty. The first is risk, such as hazardous materials, electrical equipment, or biological specimens. The second is productive struggle: sustaining challenge at the level that builds capacity without tipping into unproductive frustration that destroys motivation. Managing both distinguishes the role from classroom teaching and laboratory supervision.

    \item \textbf{Domain flexibility.} Unlike a subject teacher who specializes in one discipline, a Research Guide must be able to support inquiry across domains, or at minimum, to recognize when a student's question requires domain expertise the Guide does not have, and to connect the student with appropriate resources. The core skills of research are domain-general; the Research Guide teaches these transferable capacities.

    \item \textbf{Community building.} Research happens through social practices that a Research Guide must cultivate: peer critique of work in progress, journal-club-style engagement with the literature, and team-based projects that let novice researchers do authentic work collectively before any of them could sustain it alone. Beyond the group itself, a Research Guide establishes collaborations with external research labs, field sites, or domain practitioners so that student work connects to the broader research community. The AI for Scientific Research program at NYU (Section~\ref{sec:evidence}) was organized around both patterns.
\end{itemize}

% ============================================================================
% 5. What a Research Guide Profession Would Look Like
% ============================================================================

\section{What a Research Guide Profession Would Look Like}
\label{sec:profession}

We do not propose to build the full infrastructure of a profession in this paper. We propose the first step: recognizing and naming the role. We sketch here what the subsequent steps would entail, drawing on the Research Software Engineer precedent (Section~\ref{sec:intro}) and the NASEM report's framework.

\textbf{A standard title.} ``Research Guide'' is our proposal, constructed on the same principle as ``Research Software Engineer'': a compound that signals the two skill sets the role requires. A Research Guide is not a teacher (teachers transfer known knowledge) and not a researcher (researchers produce new knowledge); a Research Guide develops the \emph{ability} to do research. The way a mountain guide does not carry you to the summit but helps you learn to climb and navigate through unfamiliar terrain. The term has a surface overlap with library ``research guides'' (documents that orient students to databases); context disambiguates. We propose the name as a starting point; the community that forms around it will refine it.

\textbf{A training pathway: synthesizing existing fragments.} A Research Guide training program would combine pedagogical training with research methodology, producing practitioners who can both design learning experiences and guide authentic inquiry. This does not require the depth of a doctoral program. A Research Guide needs deep fluency in the \emph{process} of research (question formulation, methodology design, evidence evaluation, scientific communication) combined with the pedagogical skill to develop these capacities in others. The core competencies are domain-general; where domain-specific depth is needed, the Research Guide connects the student with appropriate specialists, just as a general practitioner refers to a cardiologist without needing to be one.

A Research Guide curriculum would not be built from scratch. Relevant pedagogical work has been accumulating for decades across several literatures, each addressing a piece of the problem:
\begin{itemize}[leftmargin=*]
    \item \textbf{Inquiry-based science education (IBSE).} A long tradition of K--12 and undergraduate pedagogy emphasizing student-led investigation; empirical studies find gains in science literacy and student confidence \cite{gormally2009}. Strong on classroom-level scaffolding; framed around science content rather than the research process itself.
    \item \textbf{Cognitive apprenticeship.} A general framework for teaching complex cognitive skills by making normally tacit expert processes visible through modeling, coaching, scaffolding, reflection, and fading \cite{collinsbrown1991}. It bridges the apprenticeship and structured-instruction models; it does not specify a research curriculum.
    \item \textbf{Guided inquiry design.} A structured process model, developed in library and information science, for scaffolding student-led inquiry from initiation through presentation. School-level and information-seeking oriented.
    \item \textbf{Research skill development frameworks.} Curricular scaffolds (for example, Willison and O'Regan's RSD) that map research facets (embarking, finding and generating, evaluating, organizing, analyzing, communicating, applying ethically) across levels of student autonomy. Strong on the assessment rubric side; less developed on the practitioner-formation side.
    \item \textbf{CURE (Course-Based Undergraduate Research Experience) pedagogy.} A growing literature on how to embed authentic research in undergraduate courses, defined around five dimensions: scientific practices, discovery, broadly relevant or important work, collaboration, and iteration \cite{auchincloss2014}. Biology-faculty surveys converge on two themes in what makes a research experience authentic: students engaging in the process of science, and questions whose answers are unknown and may not even exist \cite{spell2014}. Strong at the undergraduate course level; does not address pre-college or adult populations.
    \item \textbf{Mentor-training curricula.} The CIMER work at UW-Madison has produced evidence-based curricula in six mentoring competencies; a randomized controlled trial with 283 mentor-mentee pairs demonstrated measurable improvements in mentoring quality \cite{pfund2014}.\footnote{Center for the Improvement of Mentored Experiences in Research (CIMER); \url{https://cimerproject.org/}.} But CIMER trains existing mentors to mentor \emph{better}; it does not train new people to \emph{become} Research Guides.
\end{itemize}
These fragments are complementary, not competing. Each captures part of the job: IBSE and Guided Inquiry address classroom-level instructional design; cognitive apprenticeship addresses the transition from tacit to explicit; RSD addresses scaffolding across stages; CURE pedagogy addresses embedding research in undergraduate coursework; CIMER addresses improving the mentoring relationship itself. None of them alone produces a Research Guide, and none was designed to. A profession-level curriculum would unify them into a coherent pathway that produces practitioners fluent across modes of research, across stages of learner development, and across the two pedagogical models of apprenticeship and structured instruction.

The modes of research identified in Section~\ref{sec:modes} give this training pathway additional structure. Not every Research Guide needs mastery of all five modes from the start. A practitioner might enter the profession with deep fluency in hypothesis-driven empirical research and exploratory observation, then develop competency in theory construction or computational modeling through continuing education. The parallel to medicine is direct: all physicians share a core training, then specialize; a cardiologist is not expected to perform neurosurgery, but she understands enough general medicine to recognize when a patient needs a neurologist. Similarly, a Research Guide specializing in empirical and exploratory modes must recognize when a student's question calls for computational modeling or tool development, and either develop that competency or connect the student with a colleague who has it. The modes thus provide substance for both initial certification (demonstrating fluency in at least two or three modes) and professional development (adding modal competencies over time), making the career ladder not merely a measure of years served but of expanding capability.

The NASEM report's finding that ``mentorship is a learnable skill'' provides the logical foundation for the curriculum as a whole: if the skill is learnable, it can be taught; if it can be taught, it can be credentialed; if it can be credentialed, it can be professionalized \cite{nasem2019}.

\textbf{A career track.} Levels should reflect capability and scope, not merely years of service:

\begin{table}[ht]
\centering
\caption{Proposed Research Guide career levels with parallels in established professions.}
\label{tab:levels}
\small
\begin{tabular}{@{}llllll@{}}
\toprule
\textbf{Level} & \textbf{Research Guide} & \textbf{Sports} & \textbf{Music} & \textbf{Academia} \\
\midrule
1 & Research Guide & Assistant Coach & Studio Teacher & Adjunct / Lecturer \\[3pt]
2 & Senior Research Guide & Head Coach & Principal Teacher & Assistant Professor \\[3pt]
3 & Research Guide Director & Athletic Director & Department Chair & Associate Professor \\
\bottomrule
\end{tabular}
\end{table}

Each level expands scope from individual mentoring to institutional leadership to profession-shaping. The parallels illustrate that every field that takes talent development seriously has built a career ladder, except research education. A standardized career ladder would also enable standardized job postings: a school seeking an RG-1 and a university seeking an RG-3 Director would use the same professional vocabulary, making the role visible in job markets and allowing candidates to navigate the profession the way coaches, musicians, and academics already can. Modal specializations would sharpen this further: a biology department launching a computational research program could seek an ``RG-1, computational modeling and data-driven methods''; a museum designing a citizen science initiative could seek an ``RG-1, exploratory and tool development.'' No mode is more advanced than another; an RG-1 may enter the profession through any specialization. Today, these positions are advertised as ``science teacher with research interest'' or ``postdoc willing to mentor,'' descriptions that communicate neither the skill set required nor the professional identity offered. Standardized modal vocabulary would make the labor market clear to both employers and candidates. This structure allows practitioners to advance without leaving education, solving the current problem in which the only career progression for a talented research teacher is to leave teaching entirely.

\textbf{Institutional recognition.} The NASEM report recommends that institutions ``reward effective mentorship visibly, incorporating mentoring quality into promotion, tenure, and hiring decisions'' and that funding agencies ``require evidence-based mentorship education, mentoring compacts, and outcome assessments in grants'' \cite{nasem2019}. These recommendations would apply directly to Research Guides.

\textbf{A professional organization.} The RSE movement built a community through a combination of a manifesto, an annual survey, a conference series, and a registered professional society. A Research Guide community would benefit from the same infrastructure: a shared identity, a forum for sharing practice, and collective advocacy for recognition and funding.

% ============================================================================
% 6. The Role of AI
% ============================================================================

\section{Why Now: AI Makes Research Guides More Important, Not Less}
\label{sec:ai}

The urgency of formalizing the Research Guide profession has increased, not decreased, with the rise of AI in science.

Substantial venture capital has been invested in AI systems that automate scientific discovery: generating hypotheses, running experiments, writing papers. These tools are valuable and will accelerate discovery. But evidence is accumulating that heavy AI use shifts the character of the work: a study of 41.3 million research papers found that AI adoption narrows the topics scientists pursue and reduces engagement between them \cite{hao2026}. A similar effect has been documented at the skill level: endoscopists who used AI-assisted polyp detection became measurably worse at detecting polyps when the AI was turned off \cite{budzyn2025}.

As AI automates standardized processes across professions \cite{wef2025}, the human contribution shifts toward the skills that resist automation, such as asking the right question, evaluating whether an answer is meaningful, and exercising judgment under uncertainty. These are precisely the skills developed through authentic research training: problem-finding, tolerance for ambiguity, judgment under uncertainty, persistence through failure. Research training is not only for future researchers; it develops the capacities that every professional will need as routine cognitive work is automated. A professional who can formulate the right question will extract far more value from AI tools than one who cannot. The people who will direct AI systems, interpret their outputs, catch their errors, and decide which questions are worth asking must develop those capacities through the experience of doing research with guidance. That guidance is what Research Guides provide.

AI also has a role within the profession. AI systems designed to support research mentoring can extend the reach of human Research Guides, providing adaptive scaffolding at a scale no individual can achieve. AI and the Research Guide profession are complementary: AI provides scale; human Research Guides provide judgment, emotional support, and the modeling of intellectual dispositions that no system can replicate.

The same economic forces that make Research Guides more necessary also make the profession viable. As AI automates routine cognitive work across industries, it frees human capacity for the work AI cannot do: mentoring, developmental assessment, and the modeling of intellectual dispositions. Current investment in research mentoring is sporadic and structurally fragile: programs depend on individual champions and dissolve when those champions move on. This fragility reflects a lack of professional infrastructure, not a lack of demand. When a function has no professional name, it cannot appear as a budget line item, attract dedicated funding streams, or build institutional permanence. Naming the profession is not merely a symbolic act. It is a prerequisite for sustainable investment.

% ============================================================================
% 7. Conclusion
% ============================================================================

\section{Conclusion}
\label{sec:conclusion}

To carry a learner continuously from general exposure through serious practice, music and sport field a distinct profession at every tier: general instruction, specialist guidance for young people with demonstrated aptitude, and professional coaching. Their middle-tier professionals are the school band director and the school-team coach. Research, so far, has only the classroom science teacher and the PhD advisor; the person in between, who guides learners with aptitude or serious interest through authentic research outside the doctoral track, has no name. That is what this paper proposes to fix.

We propose the name \textbf{Research Guide} for this role: the Tier~2 profession that develops others' capacity to do research through cognitive apprenticeship, combining pedagogy, research fluency, and the developmental judgment to assess open-ended work across at least five modes of inquiry. No existing training program produces this combination; neither teacher education nor doctoral training was built for it. The name is a beginning, not a solution. Training pathways, career structures, hiring standards, developmental assessments, institutional recognition, and a professional community must all be built. They are the infrastructure a profession builds inside itself once it exists, not documents that can be imposed on it from outside.

The stakes are substantial and the timing is not incidental. Hundreds of thousands of middle and high school students already pursue authentic research each year, even more college undergraduates participate in research with a faculty member, and millions of adults engage in citizen science; each population is served today by practitioners whose judgment has no professional vocabulary in which to travel. A rubric without a profession is a document; inside a profession, it is a vocabulary. The urgency is compounded by AI: as routine cognitive work is automated across industries, the skills that authentic research develops (including problem-finding, judgment under uncertainty, and recognizing when an answer is meaningful) are precisely the skills that remain distinctively human. Evidence already shows heavy AI use narrows the topics researchers pursue and erodes underlying skill when the tool is removed. A profession whose purpose is to develop those capacities becomes more necessary, not less.

The Research Software Engineer precedent shows it has been done before, for a different invisible role: in 2012 a small group of research-software practitioners realized that their work had no name, coined one, and within thirteen years built training programs, fellowships, a professional society, and an international community of thousands. The naming was the decisive act; everything else followed from having a name. For research, that name is Research Guide.

% ============================================================================
% References
% ============================================================================

\clearpage

\noindent\textbf{Note on references.} Every reference in this paper is available in full text without institutional access or payment. While other influential work exists on relevant topics, we intentionally include only sources that any reader can obtain and read.

\bibliographystyle{paper_a}
\bibliography{paper_a_v8_0}

\section*{Author Note}

Both authors have practiced the role this paper proposes to formalize. Each trained in one of the two traditions this paper identifies as historical precedents, brought that training into the American education system, and encountered the same professional gap from different directions. Samsonau, trained in both the Soviet \emph{fizmat} tradition and American research physics, established research labs at the Princeton International School of Mathematics and Science (PRISMS) and founded AI for Scientific Research (AIfSR) at NYU, where over 100 students completed 20+ authentic research projects across seven semesters. Pearce, trained in the Nuffield approach as a Physics student and later taught this approach as an A-Level Physics teacher at Latymer Upper School. He also managed the Science \& Technology Division and directed the Senior Research Mentorship Program at Thomas Jefferson High School for Science and Technology, and as Principal of PRISMS he introduced and established the research program currently in operation there. The arguments draw on over 40 combined years of first-hand observation: similar work, performed under a different title at every institution.
\end{document}